# Organizational Fluidity and Sustainable Cooperation

Natalie S. Glance and Bernardo A. Huberman

glance@parc.xerox.com
huberman@parc.xerox.com
Dynamics of Computation Group
Xerox Palo Alto Research Center
Palo Alto, CA 94304

## Abstract

We show that fluid organizations display higher levels of cooperation than attainable by groups with either a fixed social structure or lacking one altogether. By moving within the organization, individuals cause restructurings that facilitate cooperation. Computer experiments simulating fluid organizations faced with a social dilemma reveal a myriad of complex cooperative behaviors that result from the interplay between individual strategies and structural changes. Significantly, fluid organizations can display long cycles of sustained cooperation interrupted by short bursts of defection.

March 12, 1993

# 1 Introduction

The study of social dilemmas provides insight into a central issue of social behavior: how global cooperation among individuals confronted with conflicting choices can be secured. In recent work, we have shown that cooperative behavior in a social setting can be spontaneously generated, provided that the groups are small and diverse in composition, and that their constituents have long outlooks [1, 2]. Furthermore, the emergence of cooperation takes place in an unexpected fashion, appearing suddenly and unpredictably after a long period of stasis.

In our model of ongoing collective action, intentional agents make choices that depend on their individual preferences, expectations and beliefs as well as upon incomplete knowledge of the past. Because of the nature of individuals' expectations, a strategy of conditional cooperation was shown to emerge from individually rational choices. According to this strategy, an agent will cooperate when the fraction of the group perceived as cooperating exceeds a critical threshold.

Since the critical threshold grows with group size, beyond a certain size, cooperation is no longer possible in flat, structureless groups [1, 3]. However, groups are often characterized by a distinct social structure that emerges from the pattern of interdependencies among individuals. Clustering in a social structure effectively decreases the size of the group as each individual cares most about the behavior of his/her own cluster. Accordingly, we will show how cooperation is a more likely outcome for hierarchically structured groups. In addition, isolated clusters of cooperation can survive and even trigger a cascade of cooperation throughout the rest of the organization.

The potential for cooperative solutions to social dilemmas increases further if groups allow for structural changes. In a fluid structure, the pattern of interdependencies can vary widely over time, since the sum of small local changes in the structure of a group results in broad restructurings. We find that fluid groups show much higher levels of cooperation over time. By moving within the group structure, individuals cause restructurings that enable cooperation. Computer experiments simulating fluid organizations faced with a social dilemma reveal a myriad of complex behaviors that result from the interplay between various amounts of cooperation and structural changes.

The advantages of fluidity must be balanced against possible losses of effectiveness for an organization. By effectiveness, we mean how productive a given organization is in obtaining an overall utility over time. The production function for a social good will depend on the tightness or looseness of clustering, and possibly on the stability of the social structure. The form of the production function is not known *a priori*; however, given an optimal level of clustering for a particular good, we illustrate that there is an ideal range of fluidity associated with it.

Although our results are predicated upon the assumption of hierarchical structures, we expect them to generalize to less constrained forms of social networks. Any types of structure which exhibit clustering should qualitatively behave like the hierarchies studied herein. Whether or not this is the case, the effect of the structure and fluidity of social networks on social behavior remains an interesting theoretical avenue to explore, both within and outside the study of social dilemmas.

## 2 The Topology of Organizations

Organizational theorists study group structure in order to elucidate the nature and working of the firm. In firms, there is generally an informal structure that emerges from the pattern of affective ties among participants as well as a formal one imposed from above [4]. In our study, we will be interested in the former type, the network of activities and interactions that link individuals in a group, commonly known as the sociometric structure.

A school of sociologists and social psychologists also regard the social structure of a group as elemental in understanding group behavior. The approach called social network analysis began with Barnes' [5] and Bott's [6] first attempts to use the relationship of the linkages in a network to interpret social action, and has gained momentum in the previous decade [7]. From the micro view of individual interdependencies emerges a global perspective on social structures.

Thus, organizational structure in social groups can spontaneously emerge from the pattern of interactions of group members. In contrast with its typical usage in sociology, we use the notion of group in a loose sense, defining it as the community affected by a particular problem or situation. There are a number of different general types of structural topologies: flat, or structureless, hierarchical, matrix, circular, linear, and many others. In this paper we restrict ourselves to the first two.

In addition to its topology, an organizational structure can also be described by the amount of fluidity it exhibits. Fluidity encompasses such features as how flexible the structure is, how readily individuals can locally modify structure by changing the strength of their interaction with others. The concept of fluidity also has a precedent in the social sciences; for example Srinivas and Béteille [8] state that "a network even when viewed from the standpoint of a single individual has a dynamic character. New relations are forged, and old ones discarded or modified."

Consider, as a concrete example, the social problem of limiting air pollution. This is a problem that on one level the whole world faces and must solve collectively. The common good is clean air with a minimum of pollution, not only to make living conditions better today, but, more importantly, to ensure that the world remain hospitable to life in the future. Clean air is a common good because everyone benefits from it independently of other individual's efforts to limit pollution.

In this example, the impact of pollution depends partially on relative geographical locations. That is, neglecting prevailing winds and currents, a person is more bothered when her neighbor burns his compost pile than when someone across town does the same. Similarly, the cumulative effect of everyone in town driving their cars to the popular bookstore (instead of bicycling) affects a local resident much more than someone else who lives miles away. Of course, some individuals' spheres of influence will range much further than others'. This dilution of impact with distance can be represented as a hierarchy of interactions which reveals itself in the unraveling of layer upon layer: neighborhood, town, county, state, world. The effect of one individual's actions on another depends on how many layers apart they are.

Within this hierarchical structure there is some fluidity. Although people are constrained by their resources and personal ties, they can often choose to move to a new location to escape a neighbor's radio or a textile company's fumes.

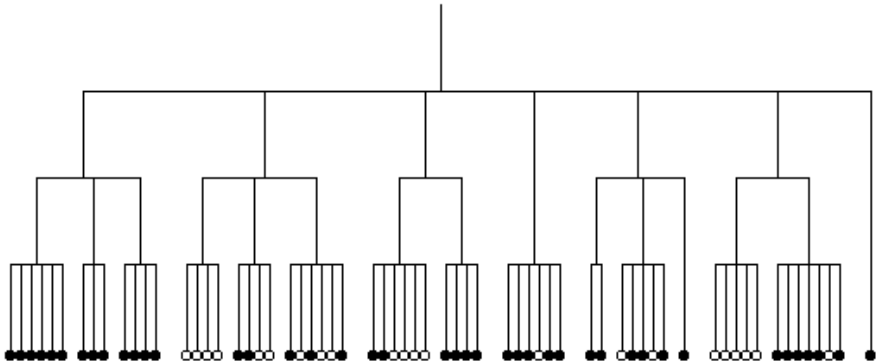

**Fig. 1.** A hierarchical organization can be visualized as a tree. The tree reveals the structure of the whole: each branching represents a subdivision of the higher level. The nodes at the lowest level represent individuals, filled circles mark cooperators, open circles, defectors. One can read directly from the tree the number of layers of the organization separating any two individuals by simply counting nodes backwards up the tree from each individual until a common ancestor is reached. This number determines the distance between two individuals. The larger the distance between them, the less their actions will affect each other.

We now quantify the notion of many levels of clustering within a group. For a group with a hierarchy of connection strengths among the members, the organizational structure can be represented graphically by a tree, as in Fig. 1. The technique of representing the interdependencies in social networks by hierarchical clustering in tree-like structures dates from the work of Hartigan [9], and more recently, Burt [10]. The level of interaction between two individuals can be read directly from the tree. Tightly clustered individuals share an ancestor node one level up in the tree. Those more loosely coupled may be connected two levels up. In general, interaction strength is indicated by the number of levels to a common ancestor node. Thus the tree gives a visual description of the amount and extent of clustering in the group. Note that this interpretation of hierarchy as a pattern of interdependencies is divorced from any notion of rank within the group.

A second more general way to describe the pattern of interactions among individual members of a group is to define the matrix of interdependencies between them. Define **A** to be the matrix of interactions; then $a_{ij}$ is the strength of interaction between individual *i* and *j*. This formalism permits generalization to many other topologies of group structure apart from hierarchical ones.

From this formalism it is easy to determine the extent of influence on any particular individual by the rest of the group. We call the cumulative influence perceived by individual *i* the rescaled quantity

$$\tilde{n}_i = \sum_j a_{ij}. \tag{1}$$

In a flat group, one in which all influences are equal, **A** would be filled with 1's, and in a group of size *n*, the rescaled size $\tilde{n}_i = n$, for all *i*. In a hierarchical structure, the components, $a_{ij}$, decrease with the number of levels separating *i* and *j*. One way to scale the strength of interaction with distance in the tree is to let

$$a_{ij} = \frac{1}{a^{d(i,j)}} \tag{2}$$

where $d(i,j)$ is the number of levels separating *i* and *j* and *a* is the scaling factor ($d(i,j) = 0$ for members of the same cluster). With this scaling, a cluster of size *a* one level distant from individual *i* is equivalent, from *i*'s point of view, to one agent in the same cluster as *i*. The rescaled size then becomes

$$\tilde{n}_i = \sum_j \frac{1}{a^{d(i,j)}}. \tag{3}$$

Along with social structure comes the notion of fluidity. In a fixed structure the pattern of interactions remains fixed in time. This need not be the case. In the example tying relative geographical positions to influence, the pattern of interactions changes slightly every time someone moves to a new locale, and more significantly when whole groups dissolve and reform. In a corporation, for example, employees may have some leeway to switch departments or even to move between regional branches. More importantly, the sum of small local changes in the structure of a group can result in broad restructurings over time.

The ease with which such moves can be accomplished reveals the amount of fluidity in the group structure. The notion of fluidity actually consists of two elements. One is how easily individuals can move within the structure, the second, how easily they can break away on their own, thereby extending the structure. These moves are restructurings in the sense that they change the pattern of interactions. Breaking away can range from moving to a secluded area, to branching off to start a new work group, to founding a new company. In many structured groups there will be costs associated with both moving within the structure or breaking away. The higher the costs the less fluid the structure will be, as individuals must anticipate a clear advantage to themselves before moving or breaking away.

## 3 Social Dilemmas

There is a long history of interest in collective action problems in political science sociology, and economics [11, 12]. Hardin coined the phrase "the tragedy of the commons" to reflect the fate of the human species if it fails to successfully resolve the social dilemma of limiting population growth [13]. Furthermore, Olson argued that the logic of collective action implies that only small groups can successfully provide themselves with a common good [14]. Others, from Smith [15] to Taylor [16, 17], have taken the problem of social dilemmas as central to the justification of the existence of the state. In economics and sociology, the study of social dilemmas sheds light on, for example, the adoption of new technologies [18] and the mobilization of political movements [19].

In a general social dilemma, a group of people attempts to obtain a common good in the absence of central authority. Each individual has two choices: either to contribute to the common good, or to shirk and free ride on the work of others. The payoffs are structured so that the incentives in the game mirror those present in social dilemmas. All individuals share equally in the common good, regardless of their actions. However, each person that cooperates increases the amount of the common good by a fixed amount, but receives only a fraction of that amount in return. Since the cost of cooperating is greater than the marginal benefit, the individual defects. Now the dilemma rears its ugly head: each individual faces the same choice; thus all defect and the common good is not produced at all. The individually rational strategy of weighing costs against benefits results in an inferior outcome — no common good is produced.

However, the logic behind the decision to cooperate or not changes when the interaction is ongoing since future expected utility gains will join present ones in influencing the rational individual's decision. In particular, individual expectations concerning the future evolution of the game can play a significant role in each member's decisions. The importance given the future depends on how long the individuals expect the interaction to last. If they expect the game to end soon, then, rationally, future expected returns should be discounted heavily with respect to known immediate returns.

On the other hand, if the interaction is likely to continue for a long time, then members may be wise to discount the future only slightly and make choices that maximize their returns on the long run. Notice that making present choices that depend on the future is rational only if, and to the extent that, a member believes its choices influence the decisions others make.

One may then ask the following questions about situations of this kind: if agents make decisions on whether or not to cooperate on the basis of imperfect information about group activity, and incorporate expectations on how their decision will affect other agents, then how will the evolution of cooperation proceed? In particular, how will the structure and fluidity of a group affect the dynamics?

In [1] we borrowed methods from statistical thermodynamics [20], a branch of physics in order to study the evolution of social cooperation. This field attempts to derive the macroscopic properties of matter (such as liquid versus solid, metal or insulator) from knowledge of the underlying interactions among the constituent atoms and molecules. In the context of social dilemmas, we adapted this methodology to study the aggregate behavior of a group composed of intentional individuals confronted with social choices. This allowed us to apply results from theoretical physics to the study of the dynamics of group cooperation.

In our mathematical treatment of the collective action problem we stated the benefits and costs to the individual associated with the two actions of cooperation and defection, *i.e.* contributing or not to the social good. The problem thus posed is referred to in the literature as the *n*-person prisoner's dilemma [21, 16, 3]. We also allowed beliefs and expectations about other individuals' actions in the future to influence each member's perception of which action, cooperation or defection, will benefit it most in the long run. Using these preferences functions, we applied the stability function formalism [22] to provide an understanding of the dynamics of cooperation in the case of groups with no organizational structure. We concluded that the emergence of cooperation among individuals can take place in a sudden and unexpected fashion. This finding turns out to be useful when examining the possibility for cooperation in fluid organizations.

## 4 Structures for Cooperation

We showed in our earlier work that there is a clear upper limit, $n^*$, to the size of a group which can support cooperation. Here we examine how this limit can be stretched by a hierarchically structured group whose members are able to move freely within the group. First, we present the limitations of flat groups and groups with fixed structures. Then we show to what extent these limitations can be overcome by fluid groups. Finally, we discuss the possible trade-off between fluidity and effectiveness in organizations.

## Flat groups

Results of Monte Carlo simulations conducted in asynchronous fashion confirm the theoretical predictions for structureless groups obtained using the stability function formalism of [22, 1]. Each individual decides whether to cooperate or defect based on the criterion that the perceived fraction cooperating, $f^c$, must be greater than a critical fraction, $f^{crit}$. Uncertainty enters since these decisions are based on perceived levels of cooperation which differ from the actual attempted amount of cooperation.

As shown in the previous work, there is a critical group size beyond which cooperation will not be sustained. There is also a range of group sizes below the critical size for which the system has two equilibrium points, one with most of the group defecting, the other with most members cooperating.

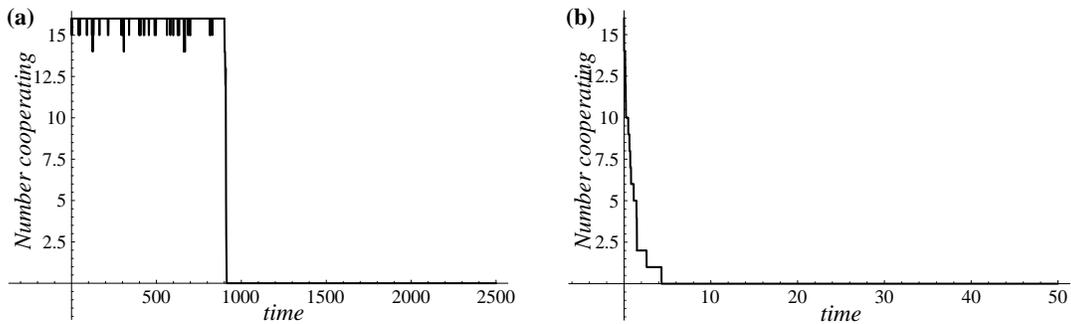

**Fig. 2.** The evolution of cooperation is shown here for a structureless group of size $n = 16$. The benefit for cooperating, $b$, is 2.5, the cost, $c$, is 1, the probability $p$ that agent cooperates successfully is 0.95, the reevaluation rate $\alpha$ is 1, and the delay $\tau$ is also 1. (a) Given these parameters, the horizon length is set at 12.0 to provide an example in which cooperation is the metastable equilibrium and defection, the global equilibrium. Indeed, the group cooperates for a very long time, over 900 time steps. The outbreak of defection is sudden and unpredictable, occurring at widely different times in numerous simulations of the same system. (b) Here the horizon length is set to 8.9; defection becomes the only equilibrium state for the system. As predicted by the theory, the group relaxes exponentially fast to the defecting state.

In Fig. 2(a), the evolution of the system is shown for a group of size 16 whose horizon length is such that the critical size for cooperation is less than 16. The system begins in the metastable overall cooperation state and remains there for a very long time. Defection is the global equilibrium, however, and eventually, the transition to defection occurs. In (b), the horizon length is such that the critical size is greater than 16. In this case, the only equilibrium point is overall defection; accordingly, the onset of defection is swift for a group initially cooperating.

## Fixed structures

Imagine now that the pattern of interactions among the 16 individuals depicts a group broken down into 4 clusters of 4 agents each. If the amount of influence one agent has on another scales down by a factor of $a=4$ for each level in the hierarchy separating two individuals, then the rescaled size of the group is $\tilde{n} = 4 + 12/4 = 7$ for all the agents (symmetric structure). If only because rescaled group size is smaller, cooperation can be sustained under more severe conditions for a hierarchical group than for a flat one.

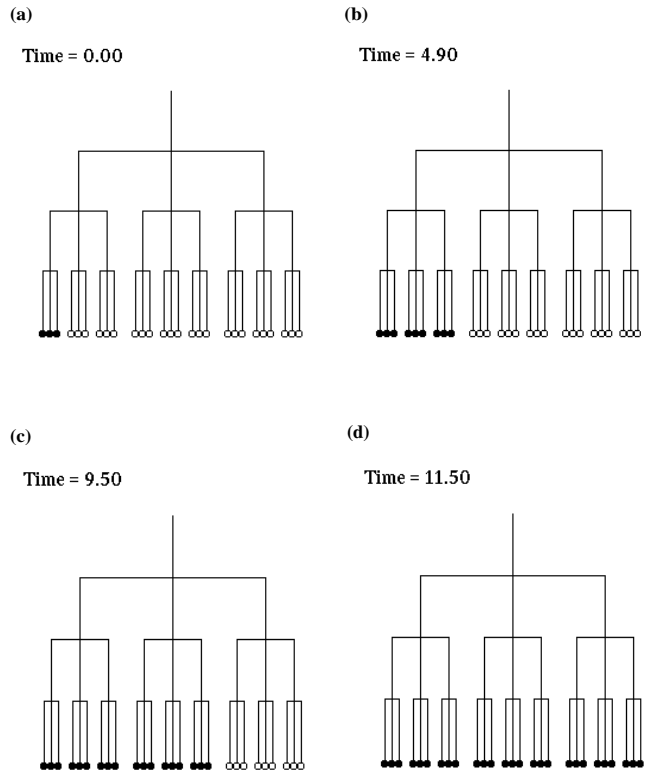

**Fig. 3.** Overall cooperation in a hierarchically structured group can be initiated by the actions of a few agents clustered together. These agents reinforce each other and at the same time can spur agents one level further removed from them to begin cooperating. In turn, this increase in cooperation can spur cooperation in agents even further removed in the structure. In this example, the structure is fixed: the three level hierarchy consists of three large clusters, each subsuming three clusters of three agents each. The parameters in this example were set to $H = 10.0$, $b = 2.5$, $c = 1$, $p = 0.97$, $\alpha = 1$, and $\tau = 0$.

More interestingly, an enclave of cooperation within the hierarchy can initiate a widespread transition to cooperation within the entire organization. The mechanism that explains this cascade to cooperation results from the clustering of agents and cannot occur in a flat group. The cascading phenomenon is more vivid in hierarchies with several layers; consider for example a group of 27 agents structured as in Fig. 3(a). In this case, the rescaled size $\tilde{n} = 3 + 6/3 + 18/3^2 = 7$ for all the agents. Filled circles at the lowest level represent cooperating agents, open circles are defecting agents. Individuals in different parts of the organization will observe widely differing amounts of cooperation. Those in the leftmost cluster see a fraction of about 3/7 cooperating while agents one level away see 1/7 cooperating and agents two levels away see a fraction 0.33/7 cooperating. So if $1/7 \lesssim f^{crit} \lesssim 3/7$, the agents in the leftmost cluster can sustain cooperation almost indefinitely among the three of them despite the fact that the rest of the organization is defecting. On the other hand, although agents one level away may be defecting initially, for certain parameter choices, cooperation may actually be the long-term stable state for those clusters in the presence of uncertainty. If these agents in clusters one level away start cooperating, they may then trigger the onset of cooperation in the clusters two levels away from the initial cooperators. Fig. 3 shows four snapshots in time of a group of 27 agents exhibiting this sequence of behavior over time.

However, as with flat groups, groups with fixed structures easily grow beyond the bounds within which cooperation is sustainable. In the case mentioned earlier of 16 individuals broken down into four clusters, when the rescaled size is larger than $\tilde{n}^* = 4.42$, the group rapidly evolves into its equilibrium state of overall defection. But even when the rescaled size is less than $\tilde{n}^*$, cooperation will be metastable in the sense that the agents may remain cooperating for a very long time but eventually a transition to overall defection will occur. When the transition finally happens, the onset of defection is observed to be very rapid, as for flat groups.

**Fluid structures**

In fluid structures, individual agents are able "move" within the organization. The way in which this might occur depends on what determines the structure, be it geographical location in the world or type of work within a company. The amount of fluidity in an organization is variable. It depends on two factors: (1) the ease with which an individual can switch between two clusters; and (2) how readily agents break away to form clusters of their own. Globally, the sum of individual moves between clusters translates into a mixing and merging of the agents. On the other hand, the local decisions made by individuals to break away and start new clusters expands the structure by decreasing the extent of clustering.

In our simulations, agents regularly reevaluate the situation. Previously, they had only one choice to make: whether to cooperate or defect. In fluid organizations, they must also evaluate how satisfied they are with their location within the structure. However, the agents are assumed to evaluate only one of these two choices at any given decision point.

Individuals make their decision to cooperate or defect according to the long-term benefit they expect to obtain, as before. In order to evaluate their position in the structure, an individual compares the long-term payoff it expects if it stays put with the long-term payoff it expects if it moves to another location, chosen randomly. In these calculations, the agent's strategy in response to the social dilemma remains the same, be it cooperation or defection. In order to determine the payoff it expects to obtain by moving, the individual must have access to the world as seen by the individual whose position it is evaluating. This additional information is not required by individuals in groups whose structure is fixed. Thus, the validity of this model of fluid organizations is limited by the extent to which this information is available.

In addition, there might be a barrier to moving, either because there is a cost associated with it, or because the agents are risk-averse. So for example, an individual might move only if it perceives the move to increase its expected payoff by a certain percentage, which we shall refer to as the moving barrier.

When evaluating its position, the individual also considers the possibility of breaking away to form a cluster of its own. The agent will do so if it perceives the payoff for either staying put or moving to be small enough that it feels it has nothing to lose by taking a chance and starting its own cluster. The agent can only break away one level at a time, so from a cluster of several agents, it may break away to form a cluster on its own one level distant from its parent cluster. The next time this same individual reevaluates its position, it can then break away an additional layer distant from its parent cluster, if no other agent has come to join it in its new cluster. In this way an agent can break away many levels from its original cluster.

How easily agents are tempted to break away determines the break away threshold. We will give these thresholds as a fraction of the maximum possible payoff over time. Higher thresholds indicate that an individual is more likely to be unsatisfied with both its present position and the alternatives and thus will tend to break away.

Computer experiments implementing this notion of fluid organizations reveal a myriad of complex behavior. Through local moves and break aways, the organization can often restructure itself to recover either from outbreaks of defection or to overcome an initial bias to defection in the group. A series of snapshots over time for a group of size 16 are shown in Fig. 4. Initially, the group is divided into four clusters of four defecting agents each — this represents an extreme condition and is thus a good test of the group's ability to overcome defection. A later snapshot shows that most of the agents have broken away on their own. This restructuring enables the global switchover to cooperation since the rescaled size is now small enough that cooperation is the global equilibrium for the system, were the structure to remain fixed. Fixed it is not, however, and over time, the now mostly cooperating agents cluster back together again, moving towards clusters where they perceive the amount of cooperation to be highest. Eventually, one or more clusters will grow too large to support cooperation indefinitely, creating the potential for an outbreak of defection. Each outbreak is quelled by a process similar to that the responsible for the initial recovery. Cycles of this type have been observed to appear frequently in the lifetime of the simulated organization.

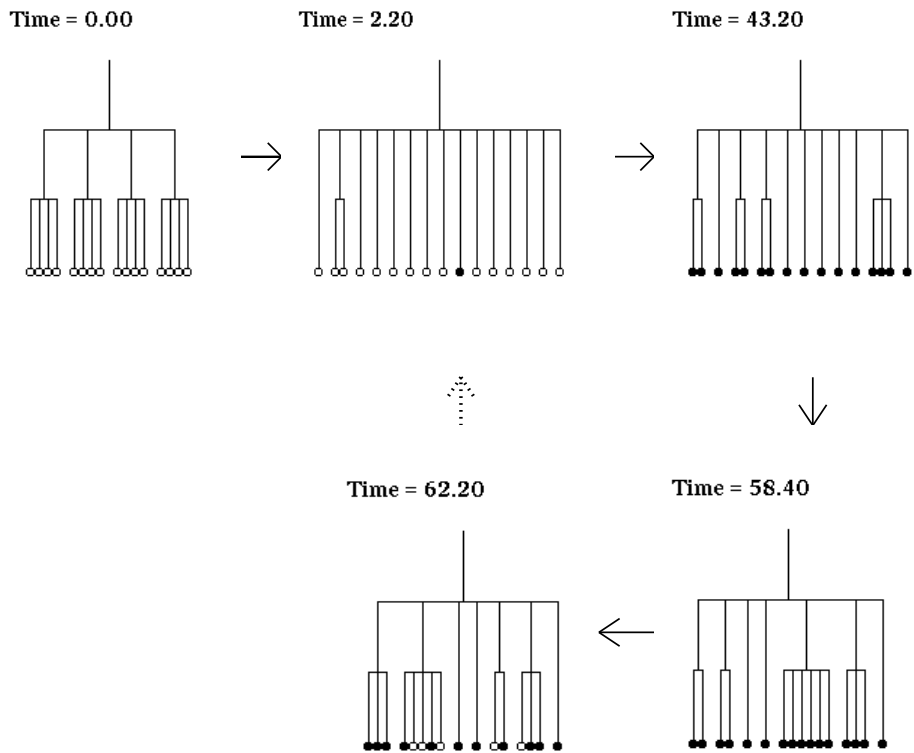

**Fig. 4.** This figure highlights the ability of fluid groups to recover from overall defection among its members. Initially all members of the group, divided into four clusters of four agents each, are defecting. The next snapshot in time shows that almost all of the agents have broken away on their own. In this dispersed structure, agents are much more likely to switch to a cooperative strategy, and indeed, by the next snapshot shown, all individuals are cooperating. Because of uncertainty, however, agents will occasionally switch between clusters. Eventually, a cluster grows large enough that a transition to defection begins within that cluster. At the same time, individuals will be moving away to escape the defectors. We see these processes happening in the fifth snapshot. At this point, more and more agents will break away on their own, and a similar cycle begins again. The parameters in this example were set to $H = 4.75$, $b = 2.5$, $c = 1$, $p = 0.9$, $\alpha = 1$, and $\tau = 0$, as in the previous figure.

**Effectiveness** *vs.* **fluidity**

In a very fluid organization, individuals break away often, thus founding new clusters, and move between clusters very readily. On the other hand, these actions rarely occur in a group with little fluidity in its structure. In the example of a fluid organization given above, the amount of fluidity of movement was set to an intermediate amount: the moving barrier was set at 15% and the break away threshold was 45% of the maximum possible payoff.

Observations of many simulations at various levels of fluidity and for differing values of horizon length point to the following conclusions. High break away thresholds mean that individuals tolerate little deviation from the maximum available payoff and often break away in search of "greener pastures." Since breaking away is the primary mechanism that allows a group to recover from bouts of defection, a greater tendency to break away is favorable in this sense. However, higher break away thresholds also cause the structure to become very dilute and disconnected. In the extreme case, agents will tend to always want to be on their own, as in the second snapshot of Fig 4. On the other hand, large moving barriers inhibit the clustering of agents and cause the structure to vary little over time. Thus large moving barriers help stabilize the structure. If the group is cooperating, its structure may then remain relatively fixed over time. Coupled with high break away thresholds, large moving barriers mean that a cooperating system may be frozen into a structure with a very small amount of clustering.

In general, the combination of ease of breaking away and difficulty of moving between clusters enable the highest levels of cooperation. However, this yields the counter-intuitive result of cooperating dis-organizations! That is, agents are all cooperating, but on their own.

The reason this seems counter-intuitive is because, thus far, the effectiveness of the organization has not been considered. By effectiveness, we mean how productive a given organization is in obtaining an overall utility over time. This is revealed by how well an organization achieves its goals. It seems reasonable to assume that each organization operates most effectively when its structure exhibits a certain amount of clustering. The ideal amount of clustering for a particular organization will depend on type of good the group is attempting to provide itself with and how this good is produced. Determining the optimal amount of clustering for a given type of organization is beyond the scope of this paper; however, we can say something about the range of fluidity that allows an organization to be most effective given an optimal level.

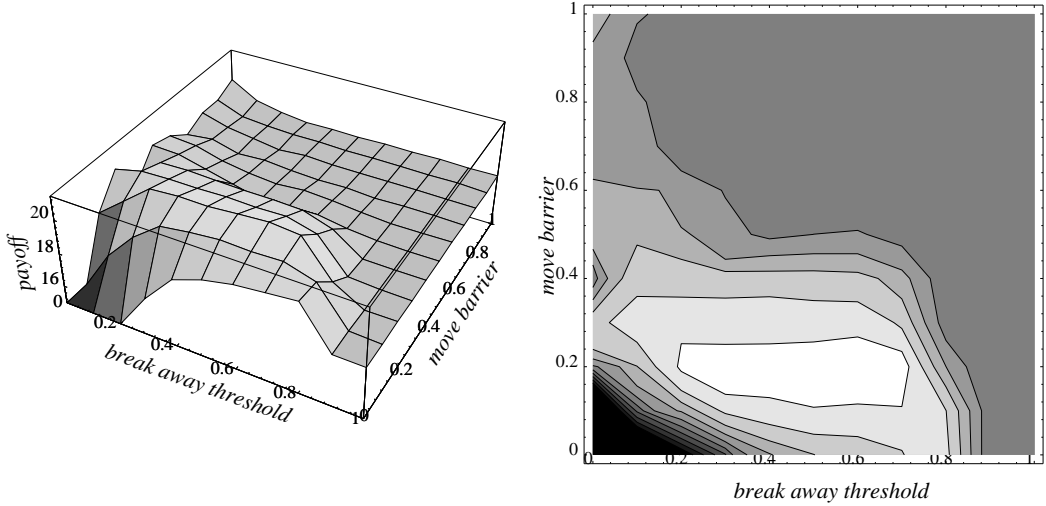

**Fig. 5.** In (a) the total actual payoff to the group, averaged over thousands of time steps, is plotted as a function of the break away threshold and move barrier. High break away thresholds and low move barriers correspond to high amounts of fluidity in the structure of the organization. The contour plot is given in (b) to highlight the gradient of the functional dependence of payoff on fluidity. In this example, the optimal amount of clustering is taken to be $\tilde{n}^{clust} = 10$, a level corresponding to the clustering of the group into two large subgroups. These results show that there is a range in the amount of fluidity at which this organization operates most effectively. The other parameters values in this example are $H = 8$, $b = 2.5$, $c = 1$, $p = 0.95$, $\alpha = 1$, and $\tau = 0$.

To factor in effectiveness, we must modify the production function of the common good to reflect increased performance at optimal clustering levels. One way to do this is to posit that the benefit to the group for a cooperative action depends on the amount of clustering. To this end, we introduce a variable

$$\tilde{n}^{clust} = \frac{1}{n} \sum_i \tilde{n}_i \quad (4)$$

that indicates the average amount of clustering within the structure and postulate that the benefits of cooperation are highest when the amount of clustering is equal to the optimal amount $\tilde{n}^{opt}$ and falls off to either side. Qualitatively, then, the benefit of an individual's cooperative action looks like

$$b' = b \exp\left[-\left(\tilde{n}^{opt} - \tilde{n}^{clust}\right)^2 / n^2\right], \quad (5)$$

where $b$ is the benefit in a flat group.

The interplay between fluidity and effectiveness is best observed when the critical rescaled size falls inside a certain regime. Within this regime, the critical size is such that high levels of clustering cause the group to be unstable to defection while low levels allow to group to recover. To measure effectiveness, we keep track of the average actual payoff over time for varying amounts of fluidity, given a fixed optimal average amount of clustering. Once again, we consider a group of 16 agents, with a hierarchical structure two levels deep. The ideal amount of clustering is set at $\tilde{n}^{clust} = 10$. Such a high level of clustering occurs when the agents break up into two large clusters. Fig. 5 shows that there is indeed a range in the amount of fluidity at which the organization operates most effectively. Systems operating within this range of fluidity may cooperate less over time than those with less fluidity, but compensate by having higher levels of clustering.

# 5 Discussion

We have shown in this paper that fluid organizations display higher levels of cooperation than attainable by groups with either a fixed social structure or lacking one altogether. In fluid organizations, individuals can easily move between groups or even strike out on their own. The sum of many such moves results in a global restructuring of the organization over time.

In an organization faced with the social dilemma of providing itself with a collective good, such ongoing incremental restructurings can provide free riders with the incentive to cooperate. That is, in a group with many free riders, individual moves will cause the structure to disperse into many small clusters. Cooperation is much more likely to emerge spontaneously within these small clusters and then spread to the rest of the organization.

The results of extensive computer experiments bear out these conclusions. In particular, fluid organizations can display long cycles of sustained cooperation interrupted by short bursts of defection. The average level of cooperation sustained over time depends on the amount of fluidity in the organization as well its breadth and extent. A point to consider, however, is that the advantages of fluidity must be balanced against a possible accompanying decrease in the effectiveness of the group.

Although our model assumes hierarchical groups faced with a social dilemma, we expect our results will generalize to less constrained forms of social networks and to more general production problems. Thus, fluidity may play a crucial role in ensuring that a large organization attain high levels of cooperation. In fact, modern firms have begun to foster organizational fluidity: a central theme in recent reorganizations of large corporations has been the creation of pathways that allow project-centered groups to rapidly form and reconfigure as circumstances demand. A recent cover story in *Fortune* magazine claims that "corporations are finally realizing the need to recognize the informal organization, free it up, and provide it the resources it needs," citing such corporate powerhouses as Apple Computer, Levi Strauss, and Xerox [23].